\documentclass[12pt]{article}

\usepackage[english]{babel}		%Franais/Anglais
\usepackage[applemac]{inputenc}	%Caractres accentus
\usepackage[T1]{fontenc}			%Csure
\usepackage{amsmath}
\usepackage{graphicx}
\usepackage[colorinlistoftodos]{todonotes}

\usepackage{amsmath}
\usepackage{amssymb}
\usepackage[retainorgcmds]{IEEEtrantools}
\usepackage{dsfont}
\usepackage[colorlinks,linkcolor=blue,citecolor=blue,urlcolor=blue]{hyperref}
\usepackage{graphicx}
\usepackage{psfrag}
\usepackage{ulem}
\usepackage{color}
\usepackage{float}
\usepackage{array}

\usepackage{amsmath,color}
\usepackage{graphicx} 
\definecolor{green}{rgb}{0.3,0.7,0.6}

\begin{document}

\title{Molding Molecular and Material Properties by Strong Light-Matter Coupling}

\maketitle

{ Branko Kolaric}$^{1,**}$, {Bjorn Maes}$^{1*}$, { Koen Clays}$^{2*}$, { Thomas Durt}$^{3*}$, {Yves Caudano}$^{4*}$\\
$^{1}$Micro- and Nanophotonic Materials Group, University of Mons, Place du Parc 20, 7000 Mons, Belgium\\
$^{2}$Department of Chemistry, KU Leuven, Celestijnenlaan 200D, 3001 Heverlee, Belgium and Department of Physics and Astronomy, Washington State University, Pullman, WA, USA\\
$^{3}$Aix Marseille Univ, CNRS, Centrale Marseille, Institut Fresnel, F-13013 Marseille, France\\
$^{4}$Department of Physics, University of Namur, Rue de Bruxelles 61, 5000 Namur, Belgium\\

$^{**}$branko.kolaric@umons.ac.be
\footnote{All authors are senior scientists and contributed equally (share $*$) for this review. Additionally, BK has been responsible for the correspondance with the editorial office acting like 'primus inter pares'.}

\begin{abstract}
When atoms come together and bond, we call these new states molecules and their properties determine many aspects of our daily life. Strangely enough, it is conceivable for light and molecules to bond, creating new hybrid light-matter states with far-reaching consequences for these {\it strongly coupled} materials. Even stranger, there is no `real' light needed to obtain the effects, it simply appears from the vacuum, creating `something from nothing'. Surprisingly, the setup required to create these materials has become moderately straightforward. In its simplest form, one only needs to put a strongly absorbing material at the appropriate place between two mirrors, and quantum magic can appear. Only recently has it been discovered that strong coupling can affect a host of significant effects at a material and molecular level, which were thought to be independent of the `light' environment: phase transitions, conductivity, chemical reactions, etc. This review addresses the fundamentals of this opportunity:  the quantum mechanical foundations, the relevant plasmonic and photonic structures, and a description of the various applications, connecting materials chemistry with quantum information, nonlinear optics and chemical reactivity. Ultimately, revealing the interplay between light and matter in this new regime opens attractive avenues for many applications in the material, chemical, quantum mechanical and biological realms.
\end{abstract}

\section {Introduction}

All matter around us is made by combining atoms, the constitutive building blocks of our universe, in different ways \cite{Atkins}. Many groups of atoms are then identified to exist together as one species having characteristic properties \cite{Atkins,Fulz}. Atoms in different chemical species or material entities (such as molecules, ions, and crystals) are held together by attractive forces, called a chemical bond. The key dogma of modern physical and chemical sciences is that some intrinsic properties are transferable from one molecule to another because they are attributed to atoms and functional groups, independently of their environment. Furthermore, current material science is developed around the idea that molding chemical bonds is the essential condition for tuning physical properties and chemical reactivity. 
Here in this short review, we shall highlight new and attractive findings from the field of strong coupling (SC) that allow modification of material properties without altering their chemical structure in a classical way. 

Very recently it was pointed out that SC could affect a host of significant phenomena at a material and molecular level, which were thought to be independent of the light environment: phase transitions, conductivity, chemical reactions, etc. \cite{AccountsChem49-2016-2403-Ebbesen} It should be stressed that the SC regime is achieved even without the need for real light; it simply appears from the vacuum, creating something from seemingly nothing \cite{Saiko}.
Beyond a fundamental study of strong coupling in quantum optics \cite{NatPhot8-2014-685-Lukin, PRA94-2016-063825-Carreno}, a long-term research objective was to evidence SC of light and matter \cite{AccountsChem49-2016-2403-Ebbesen} in an experiment performed in ambient conditions. This effort succeeded recently, using various plasmonic and photonic structures, with broad Rabi splitting \cite{AccountsChem49-2016-2403-Ebbesen,HarocheRaimond}. A hybridisation model can explain the essence of strong coupling  in the language of physical-chemists. The strong interaction between the material entity and light creates new hybrid light-matter states that have significantly different energy levels from those of the material entity and of the optical system individually (Fig. 1).
 
 \begin{figure}\label{figure1}
\includegraphics[width=\columnwidth]{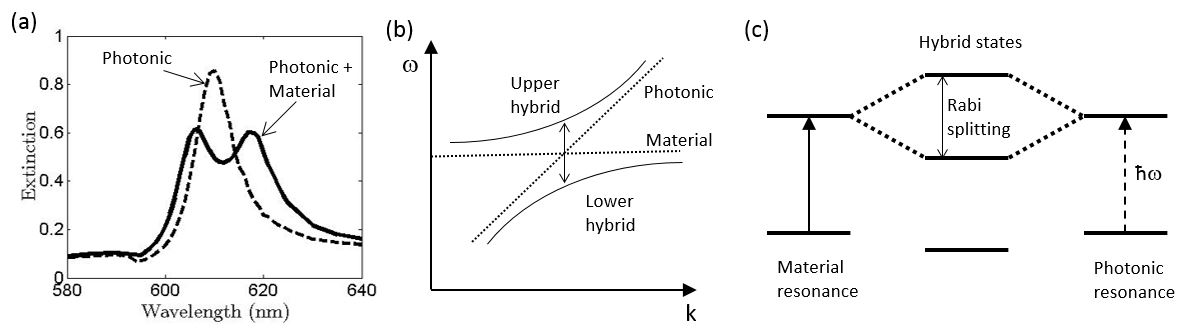}
\caption{\footnotesize \itshape (a) Photonic resonance of a plasmonic array (dashed) and peak splitting (solid) when a material is introduced on top of the array. (b) Avoided crossing: the separate modes (dotted) strongly shift upon SC (solid). (c) A material resonance (left) overlaps with a photonic mode (right) leading to two hybrid dressed states (middle), with energy difference called the Rabi splitting.}
\label{one}
\end{figure}
 
The signature of the strong light-matter coupling is typically detected through the changes in the excitation spectrum (electronic or vibrational) of the coupled system.  A quasiparticle called polariton emerges as a new hybrid entity. It offers the possibility to mold chemical and material properties in an unprecedented way.  Quantum mechanically,  polaritons are also termed dressed states. These hybrid states oscillate between their light and matter components, cycling through absorption and emission instead of exhibiting a typical decay. The unique features of these hybrid states, i.e. polaritons, are at the heart of this review and will be described in the following sections. The first step that leads to harvesting of polaritons and to the creation of new `meta physics and chemistry' beyond the chemical bond paradigm is to design various plasmonic and photonic structures that permit the strong coupling between matter and light.

In this review, we first describe the quantum mechanical aspect of strong coupling and the fundamental difference between the weak and the strong coupling regimes, in particular the influence of the coupling on emission properties. After that, we shortly review plasmonic and photonic structures used to achieve strong coupling \cite{AccountsChem49-2016-2403-Ebbesen, Kolaric}. In the end, we review applications of strong coupling to mold materials and molecular properties and highlight potential future research.

\section{Quantum mechanical aspects of strong coupling}

We usually conceive the constituents of matter as independent of their environment. This idea is central to the reductionist approach to science, which analyses systems on the basis of their elements. The identification of these elements stems from the elements' invariant properties when they are isolated. When combined, matter constituents couple, their properties are affected, and new properties specific to the system emerge. The classical hypothesis is that the global properties of the system are adequately described from their constituents and couplings. However, this conception is challenged by quantum mechanics. Indeed, on the one hand, the quantum state of a system can be precisely defined although there exists no specification of the state of its constituents. These characteristics is at the heart of the notion of entanglement in a composite quantum system. On the other hand, phenomena such as the spontaneous emission of light by an isolated molecule or atom occur because the properties of matter cannot be defined independently of their environment, be it vacuum. As a result, the radiative deexcitation that occurs in the fluorescence process is a property which, although usually associated with matter, in reality, is also inextricably linked to the electromagnetic fluctuations of the quantum vacuum \cite{HarocheRaimond}.

Most often, the coupling between matter and the electromagnetic field is weak. As a result, in practice, it is possible to consider matter and its electromagnetic environment as separate entities which simply exchange energy. These exchanges are evaluated using perturbation theory: the quantum calculation of the spontaneous emission rate uses thus Fermi's golden rule\footnote{According to Fermi's Golden Rule, the inverse of the lifetime of an initially excited state coupled to a continuum of final states is equal \cite{Joachain} to ${2\pi\over \hbar} g^2\rho(E)$, where $g$ is the modulus of the matrix element measuring the coupling between the initial and the final states (averaged over the final states), while $\rho(E)$ is the number of final states by unit of energy,(with the same energy as the initial excited state) corresponds to the DOS ; here we consider a Bohrian transition between an excited atomic state of energy $E_e$ and an atomic ground state of energy $E_g$ accompanied by the emission of one photon. Then, energy conservation imposes  that the frequency $\nu$ of the emitted photon is equal to $\omega_e=(E_e-E_g)/h$.}. Fermi's rule shows that the spontaneous emission rate depends on the density of states (DOS) of the coupled electromagnetic modes. Therefore, a change in the electromagnetic DOS can increase or reduce spontaneous emission. This is the Purcell effect\footnote{In a box of volume $V$, $g^2$ is equal, up to constants \cite{Joachain}, to $\nu^3\cdot d^2/V$,({\it d}-dipole moment) and $\rho(E)$ has been shown by Rayleigh and Jeans in the framework of classical electrodynamics to be equal, in the limit of $V$ going to infinity, to $8\pi\nu^2\cdot V/h$}. Denoting $F_P$ the Purcell factor, which is equal to the ratio between the lifetime in the vacuum (seen as an infinite box) and the lifetime in the cavity, we find that, up to constants, $F_P=1/(\nu^2\cdot V\cdot \gamma_c)$ as was shown by Purcell in 1946. $F_P$ depends explicitly on the volume $V$, as well as on the resonance frequency $\nu$ and the quality factor $Q=\nu/\gamma_c$ of the cavity. Thus, the environment molds the properties of the atom trapped in the cavity: the fluorescence of a quantum emitter is modified close to a reflecting surface or inside a cavity. In a cavity for instance, the coupling acts even when there is no photon in the cavity, due to quantum fluctuations. The coupling strength between the atom and Maxwell fields is characterized by a coupling factor $g$ proportional to $E_0$, the effective electric field of the empty cavity: $E_0=\sqrt{\hbar \omega_c/(2 \epsilon_0 V_c)}$, where $V_c$ is the cavity volume and $\omega_c$ is the cavity mode frequency. 

Note that $g=\Omega_0/2$, where $\Omega_0$ is called the vacuum Rabi frequency. Quantum electrodynamics predicts other phenomena as well, such as the Lamb shift: the energy of the $^2S_{1/2}$ and $^2 P_{1/2}$ orbitals of the hydrogen atom split due to the electron's interaction with the cloud of virtual photons associated to the quantum vacuum fluctuations. We see that the properties of matter cannot be dissociated from the quantum vacuum. 

Contrary to what can sometimes be read in the literature, even when the coupling remains weak, the states of matter and of the electromagnetic field cannot be defined independently, as the two subsystems of a whole. In quantum language, the system states are not factorable. For instance, the state of a system composed by a decaying atom and its surrounding Maxwell field in free space can  be written (at zero temperature and at the Rotating Wave Approximation) in the form $\alpha\lvert e\rangle_A\lvert 0\rangle_L+\beta \lvert g\rangle_A\lvert 1\rangle_L$. In this expression, $\lvert e\rangle_A$ and $\lvert g\rangle_A$ represent the excited and ground states of the atom, respectively, $\lvert 1\rangle_L$ is a single-photon state and $\lvert 0\rangle_L$ is the vacuum state, while $\alpha$ and $\beta$ are complex amplitudes. In particular, at the so-called Wigner-Weiskopff approximation, which is valid in the weak coupling regime, one finds that if at initial time $t_0=0$  one prepares an atom in the excited state, with no photon in the environment (this means that $\alpha(t=0)=1$ and $\beta(t=0)=0$), then for later times $t>0$, $\lvert \alpha(t)\lvert ^2=e^{-\Gamma \cdot t}$ and $\lvert \beta\lvert =\sqrt{1-\lvert \alpha(t)\lvert ^2}$. Entanglement is thus maximal at time $t=\ln 2/\Gamma$ (then $\lvert \alpha(t)\lvert ^2=\lvert \beta(t)\lvert ^2=1/2$). What happens in this case however is the following: the emitted photon irreversibly radiates to infinity, and from this point of view it is decoupled from the atomic source. For these reasons, it is also impossible for all practical purposes to observe interferences between the states $\lvert e\rangle_A\lvert 0\rangle_L$ and $\lvert g\rangle_A\lvert 1\rangle_L$, so that everything happens as if the full state of the system was an incoherent superposition (mixture) of these two factorable states of the type $\lvert \alpha(t)\lvert ^2\lvert e\rangle_A\lvert 0\rangle_L\langle e\lvert _A\langle0\lvert _L+\lvert \beta(t)\lvert ^2\lvert g\rangle_A\lvert 1\rangle_L\langle g\lvert _A\langle1\lvert _L.$ Considered from this perspective, the atom and the field do not ``speak'' much together and preserve their identity throughout the interaction.

The strong coupling regime, the object of this review, is remarkable, not only because it does not allow a clear distinction between the states of matter and of the field, which is also the case, strictly speaking, in the weak coupling regime as discussed above. What is remarkable in the strong coupling regime is that the emitted photon stays localized in the vicinity of the atom (this is why cavities are most often necessary for obtaining the strong coupling regime) and keeps on interacting with it. In the strong coupling regime, reversible coherent transfers occur between the atom and the Maxwell field, the so-called Rabi oscillations. As we shall discuss in the next section, in the strong coupling regime, several Rabi oscillations have time to take place before dissipation washes them away. For short times it is then consistent to neglect losses (this is the so-called Jaynes-Cummings approximation), in which case the stationary matter-field states provide a good approximation to quasi-stationary states observed during experiments. These (quasi) stationary states of strongly coupled matter-light were shown by Jaynes and Cummings to constitute a non-factorable coherent quantum superposition of the matter and field excitations. These states are entangled and are called dressed states or polaritons, according to whether we consider excitations of isolated emitters or of a collection of them. These states posses a matter component that is inseparable from the electromagnetic field. Strong coupling affects deeply the properties of matter, opening numerous fundamental and technological opportunities.

\subsection{Weak to strong coupling transition}

\subsubsection{Discrete version of the Quantum Electrodynamics (QED) Fermi Golden Rule}

\begin{figure}[h]
  \begin{center}
   \includegraphics[scale=0.75]{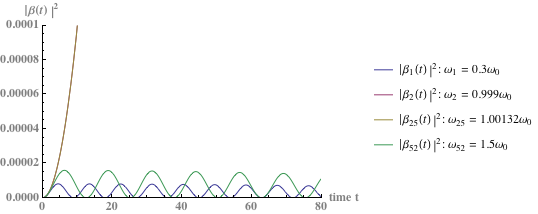}
  \end{center}
  \caption{\footnotesize \itshape Excitation (for a system with 50 modes) of two modes with frequencies within $\Delta\omega=5\cdot10^{-3}$ of $\omega_0$ (in red and yellow) to which we added two far-away mode frequencies $\omega_1<\omega_0$ (blue) and $\omega_{52}>\omega_0$ (green). The red and yellow curves overlap on the figure as a result of energy conservation.}
  \label{fig:ActiveSmall}
\end{figure}

In order to grasp the fundamental differences between weak and strong coupling, it is useful \cite{goessens,DDGB} to consider a  {\bf discrete version of Fermi Golden Rule}. To do so, let us consider an excited state coupled to a finite number $N$ of final states. The evolution is then described by a matrix of the type
\begin{equation}
\mathrm{i}\hbar\frac{\mathrm{d}}{\mathrm{d}t}
\begin{pmatrix}
\alpha(t) \\
\beta_1(t) \\
\vdots \\
\beta_N(t)
\end{pmatrix} =\hbar \begin{pmatrix}
\omega_0 & \lambda'^\star(\omega_1) & \ldots & \lambda'^\star(\omega_n) \\
\lambda'(\omega_1) & \omega_1 & & \\
\vdots & & \ddots & \text{\huge0}\\
\lambda'(\omega_N) & \text{\huge0}& & \omega_N
\end{pmatrix}  \cdot \begin{pmatrix}
\alpha(t) \\
\beta_1(t) \\
\vdots \\
\beta_N(t)
\end{pmatrix},
\end{equation}
where the excited state is characterized by the complex amplitude $\alpha$ and the final states by the $N$ amplitudes $\beta_i$. The square matrix, called $M_{(i,j)}$ in the following, was diagonalised using \emph{Mathematica}
%\begin{equation}
%M_{(i,j)}=\begin{pmatrix}
%\omega_0 & \lambda'^\star(\omega_1) & \ldots & \lambda'^\star(\omega_n) \\
%\lambda'(\omega_1) & \omega_1 & & \\
%\vdots & & \ddots & \text{\huge0}\\
%\lambda'(\omega_N) & \text{\huge0}& & \omega_N
%\end{pmatrix} 
%\end{equation}
%where for all $\omega_i$, $\lambda'(\omega_i)\equiv\lambda(\omega_i)/\hbar$ 
and we eventually took $\lambda(\omega_i)=\lambda$ for all $\omega_i$. The cavity frequency $\omega_0$ was was set to 1 and the outer frequencies $\omega_j$ were taken to belong to the spectrum
\begin{equation}
k_n = n \frac{\pi}{L} \equiv nk_1 \qquad \text{where} \quad n\in\left[1,2,3,\dots,N\right],
\end{equation}
of the infinite square well. Namely,
$\omega_j = j\omega_1$ with $j\in\left[1,2,3,\dots,N\right]$,
which should always be understood in units of $\omega_0$. The program finally gives back the eigenvalues $\kappa_i$ and eigenvectors $\vert \kappa_i \rangle$ of $M_{(i,j)}$, that can be used to solve for $\alpha(t)$ and $\beta_j(t)$ using the general expression of their solution
\begin{equation}
\begin{pmatrix}
\alpha(t) \\
\beta_1(t) \\
\vdots \\
\beta_N(t)
\end{pmatrix} = \sum_{j=1}^{N+1}c_j \vert \kappa_j\rangle \mathrm{e}^{-\mathrm{i}\kappa_j t}.
\end{equation}
The solutions for $\alpha(t)$ and $\beta_i(t)$ separately are then found by taking
 \begin{equation}
  \left\{ \begin{array}{lll}
  \alpha(t) & = & \sum_{j=1}^{N+1}c_j \langle 1_\text{in}\vert \kappa_j\rangle \mathrm{e}^{-\mathrm{i}\kappa_j t}\\
  \beta_i(t) & = &  \sum_{j=1}^{N+1}c_j \langle 1_\text{out} ^{(i)}\vert \kappa_j\rangle \mathrm{e}^{-\mathrm{i}\kappa_j t}
  \end{array} \right.
\label{eq:incond}
  \end{equation}
with initial conditions
  \begin{equation}
  \left\{ \begin{array}{lll}
  \alpha(t=0) & = & \sum_{j=1}^{N+1}c_j \langle 1_\text{in}\vert \kappa_j\rangle = 1\\
  \beta_i(t=0) & = &  \sum_{j=1}^{N+1}c_j \langle 1_\text{out} ^{(i)}\vert \kappa_j\rangle = 0
    \end{array} \right.
\label{eq:incond}
  \end{equation}
where $\lvert 1_\text{in} \rangle$ is the excited state and the $\lvert 1_\text{out} ^{(i)}\rangle$ corresponds to the $i^{th}$ final state. Our numerical results validated the property of energy conservation, which is one of the predictions of the perturbative development {\it \`a la} Fermi: as can be seen from figure \ref{fig:ActiveSmall} \cite{goessens,DDGB}, only modes fulfilling energy conservation significatively respond to the coupling.

 \subsubsection{Beyond Fermi predictions: the W-W model and exponential decay}
 
\begin{figure}[h!]
  \begin{center}
\begin{tabular}{>{\centering\arraybackslash}m{0.4\textwidth} >{\centering\arraybackslash}m{0.4\textwidth}}\includegraphics[height=0.4\textheight]{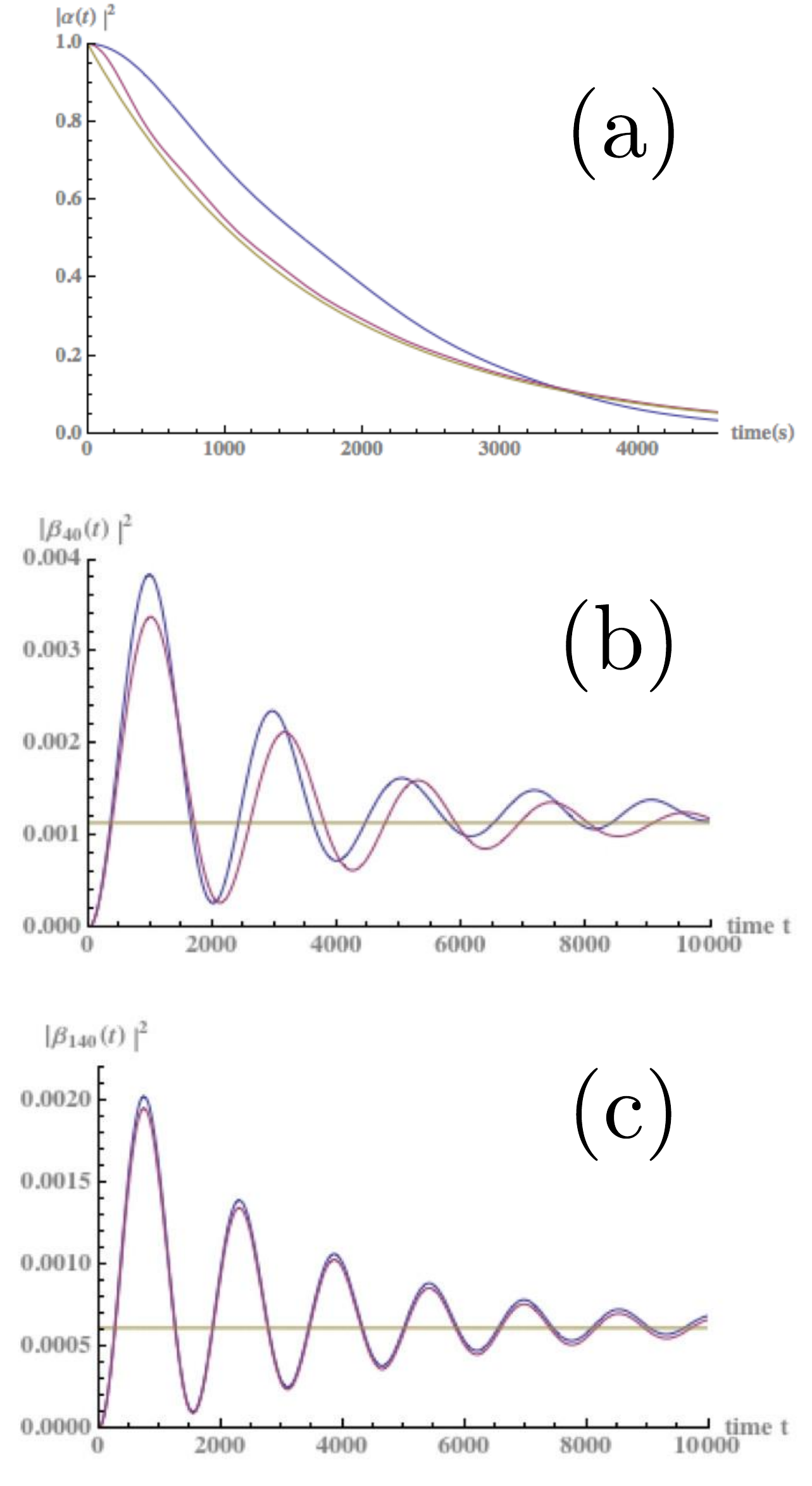} & \includegraphics[width=0.4\textwidth]{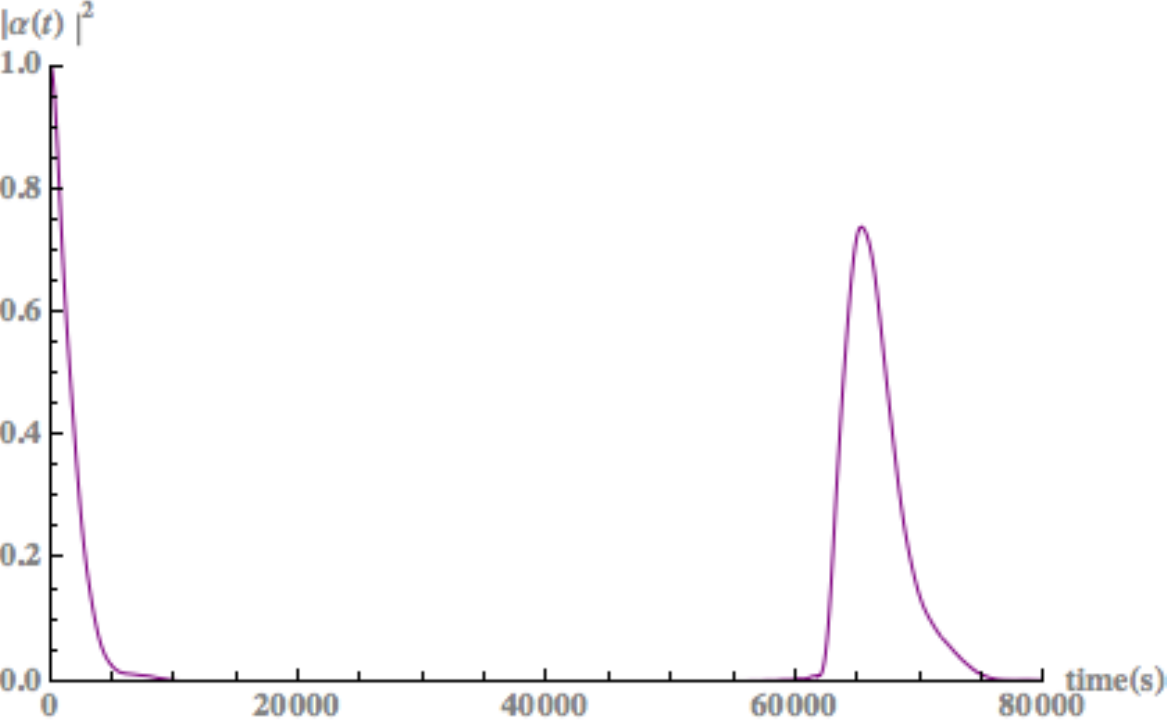}  \end{tabular}
  \end{center}
  \caption{\footnotesize \itshape {\bf Left:} a: decay of $\vert\alpha(t)\vert^2$ with $\Delta N=50$ (blue) and $\Delta N=200$ (red) modes against the theoretical expected behaviour (W-W) in yellow; Left b(c): probability amplitude $\vert\beta_{40}(t)\vert^2$ ($\vert\beta_{140}(t)\vert^2$) simulated with $\Delta N=40$ ($\Delta N=140$) modes in blue versus the W-W prediction (proportional to $ \lvert {(e^{-i(\omega-\omega_{e})t}e^{-\Gamma t}-1)\over (\omega-\omega_{e}+i\Gamma/2)}\lvert ^2$) in red and the (Lorentzian) saturation value (proportional to the Lorentz distribution in $1/((\omega-\omega_{e})^2+(\Gamma/2)^2)$) in yellow, reached for relatively long times ($t \gg 1/\Gamma$).
 {\bf Right:} appearance of a Poincar\'e recurrence for very long times.}
  \label{tabula}
   
\end{figure}

The presented numerical results scale perfectly with predictions derived in the framework of the Wigner-Weiskopff (W-W) approach, which goes beyond the Fermi approximation which is valid only for short times. The  W-W approach establishes that for longer times, the survival probability decays exponentially, while the statistical distribution of the energies of final states is a Lorentzian distribution, centered around the Bohr frequency\footnote{Fermi golden Rule is valid only when $\Gamma\cdot t $ is small. For larger times, Wigner and Weisskopf \cite{W-W} showed that (1) in good approximation the probability of survival of the initial state at time $t$ is equal to $e^{-\Gamma\cdot t}$, (2) that the probability of excitation of a final state of energy $\hbar\omega$ is equal to $\lvert <f\lvert \Delta H\lvert i>\lvert ^2\cdot \lvert {(e^{-i(\omega-\omega_{e})t}e^{-\Gamma t}-1)\over (\omega-\omega_{e}+i\Gamma/2)}\lvert ^2$ which (3) exhibits Lorentzian saturation for large times (see figure \ref{tabula}).}. This can be seen from the left part of figure \ref{tabula} \cite{goessens}.

\subsubsection{Weak and strong coupling versus Poincar\'e recurrence time}

Finally, our toy-model allowed us to study Poincar\'e recurrence time and reversibility in time of the dynamics of the system. As is well-known, the dynamics of discrete systems is (quasi)periodic. The Poincar\'e recurrence time is thus finite.
From the right part of figure \ref{tabula}, one can check \cite{goessens} that this is well so in our case.

\begin{figure}[tp!]
  \begin{center}
   \includegraphics[scale=0.5]{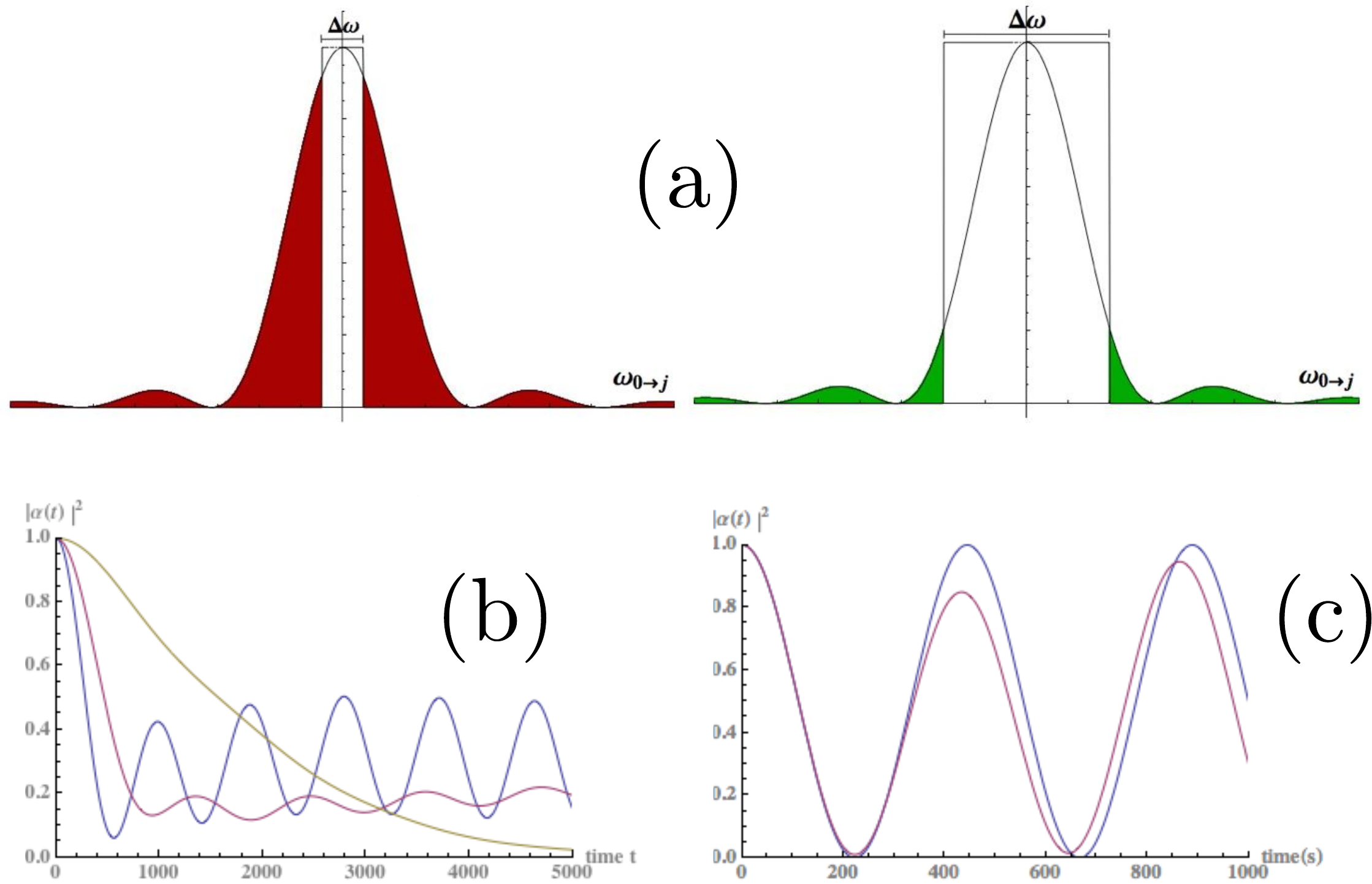}
  \end{center}
  \caption{\footnotesize \itshape a-Left: narrow bandwidth for the final modes-strong coupling regime; a-Right: large bandwidth for the final modes-weak coupling regime; b: illustration (based on the simulation of a system with $\Delta N=50 $ modes) of the strong to weak (Fermi) regime transitions by fulfilling better and better the condition $\Delta\omega \sim 2\pi\Gamma$, from blue (with $\delta \omega \approx \Gamma/2$), via red (with $\delta \omega\approx \Gamma$), to yellow (with $\delta \omega\approx 10\cdot \Gamma$); c: in red: reversible Poincar\'e recurrences of a system with a narrow bandwidth for the final modes (as in a-Left); in blue: $\mathrm{sin}^2$ modulation corresponding to an infinitely narrow distribution of final modes.}
  \label{FIG3}
\end{figure}

We noticed however (see figure  \ref{FIG3} b) that the decay times are exponentially distributed (which is the main prediction in the weak coupling regime, in accordance with Fermi and W-W treatments) if and only if the bandwidth of the final states is at least equal to $1/\Gamma$ (in agreement with the time-energy uncertainty relation). If, on the contrary, we couple the excited state to finite states having a very small frequency range $\Delta\omega$ around the Bohr frequency $\omega_e$ (so that they act as one single mode) the Poincar\'e recurrence occurs very rapidly. This can be seen from figure  \ref{FIG3} c where we compare the perfect $\mathrm{sin}^2$-shape (predicted for a conservative system {\it \`a la} Jaynes-Cummings) with the survival probability numerically predicted thanks to our toy-model.

It is worth noting that in the case corresponding to figure  \ref{FIG3} c, the system presents many analogies with  a system composed of two coupled oscillators which periodically transfer energy to each other. In the quantum situation, this exchange consists of the exchange of an elementary quantum of energy (photon). Even in this case however, the analogy with classical systems is limited, due to the absence of entanglement in the classical description. Serge Haroche and coworkers managed for instance to entangle two atoms passing successively through a cavity QED by entangling the atom with the light in the cavity in a first time, transferring afterwards this entanglement to the second atom passing through the cavity \cite{HarocheRaimond}. These properties and operations have no classical analog \cite{DV}.

\subsection{Strong coupling properties}
The quantum modeling of strong coupling relies on cavity quantum electrodynamics. The system's Hamiltonian contains two terms describing matter and radiation independently, and a third term representing their coupling through dipolar interaction. In order to grasp the typical characteristics of the strong coupling regime, we describe here the behavior of a two-level quantum system placed in a cavity (Jaynes-Cummings Hamiltonian) \cite{MicroCav-book-Kavokin}, \cite{EPJST203-2012-162-Bina}.

Let us denote $\hbar \omega_{e}$ the excitation energy of the two-level system and $d$ its transition dipole. In an empty cavity, the dipolar interaction is equal to the product $-d E_0$ between the dipole moment and the effective cavity electric field (in absence of photon).  As already mentioned, the electric field of the empty cavity is $E_0=\sqrt{\hbar \omega_c/(2 \epsilon_0 V_c)}$, where $V_c$ is the cavity volume and $\omega_c$ is the cavity mode frequency. The coupling strength is characterized by $g=d E_0/ \hbar=\Omega_0/2$, with $\Omega_0$ the aforementioned vacuum Rabi frequency.

When dissipative processes can be neglected (see previous section for an estimate of the dissipation in terms of Poincar\'e recurrence times), the new eigenenergies of the coupled system are \cite{EPJST203-2012-162-Bina}
\begin{equation}
E_\pm=\hbar \omega_c\pm\frac{1}{2}\hbar \sqrt{\delta^2+\Omega_0^2}
\end{equation}
where the detuning parameter $\delta=\omega_c-\omega_e$ describes the departure from resonance. An energy gap opens at resonance ($\delta=0$): then, the energy difference $\hbar \Omega$ between the two modes is minimal and equal to $\hbar \Omega_0$. The corresponding eigenstates are given by
\begin{equation}
\begin{aligned}
\lvert + \rangle =\sin \theta\ \lvert 0_c,e \rangle+ \cos \theta\ \lvert 1_c,g \rangle \\
\lvert - \rangle  =\cos \theta\ \lvert 0_c,e \rangle- \sin \theta\ \lvert 1_c,g \rangle\\
\end{aligned}
\end{equation}
where $\lvert 0_c,e \rangle$  represents the two-level system in its excited state with no photon in the cavity, while $\lvert 1_c,g \rangle$ describes the existence of one photon in the cavity when the two-level system is in its ground state. The parameter $\theta$ defines the entanglement strength between the two situations ($\tan 2\theta=\Omega/\delta$). At resonance, entanglement is maximal ($\theta=\pi/4$) and the dressed states have matter and field components of equal weight. The separation between two distinct matter-radiation subsystems is impossible. Far from resonance ($\theta\approx0$ and $\theta\approx\pi/2$), these states exhibit dominant matter or electromagnetic components. The polaritonic states inherit a dispersion relation from their electromagnetic component. Let us note that the pair of eigenenergies $E_\pm$ and eigenstates $\lvert\pm\rangle$ are only the first of an infinite ladder of pairs. Similar pairs exist for the quantum superposition of states  $\lvert n_c,e \rangle$ and  $\lvert n_c+1,g \rangle$ involving the presence of $n_c$ or $n_c+1$ photons in the cavity, respectively. Their energy gap at resonance is given by $\hbar \Omega_0 \sqrt{n_c+1}$.

The criterium to enter the strong coupling regime is that the Rabi frequency $\Omega_0$ must be significantly  larger than the widths $\gamma_c$ and  $\gamma_e$ related to the cavity mode and matter excitation lifetimes, respectively. In other words, the coupling must dominate over dissipative processes. This condition is required to preserve quantum coherence, as the Rabi frequency describes the rate of coherent energy conversions between matter and the radiation field in the dressed states. Another interpretation of the strong coupling condition, commonly found in the literature, is that the Lamb shift induced by the coupling is larger than the natural bandwidth of the subsystems in absence of coupling. Close to resonance, the Lamb shift is of the order of the Rabi vacuum frequency $\Omega_0$, and we recover the aforementioned condition of strong coupling  $\Omega_0 \gg \gamma_{e}$,  $\Omega_0 \gg \gamma_{c}$. As shown in a previous section, this condition can as well be interpreted in terms of Poincar\'e recurrences, in which case strong coupling occurs whenever losses are too weak to inhibit the occurence of Rabi oscillations.  It is worth noting that when an atom is put in a cavity, it can be shown \cite{Dutra} that, in the weak coupling regime, the lifetime of the dipolar excitation is equal to $\gamma_{c}/4g^2$ in accordance with Purcell prediction. Recasting the weak coupling condition $\gamma_{c}/g\gg 1$ in the form $\gamma_c^2/g^2 \gg1$, we find \cite{Haroche} that the weak coupling condition also implies that the lifetime of the cavity $1/\gamma_c$ is quite shorter than the decay time of the atom $\gamma_{c}/4g^2$ in the cavity: $1/\gamma_c \ll \gamma_{c}/4g^2$.

As a consequence of the relation  $g=d E_0/ \hbar=d \sqrt{\omega_c/(2 \epsilon_0 V_c)/\hbar}$, to increase the coupling $g$ (or, equivalently, to increase the Rabi pulsation $\Omega_0$), one needs to select quantum emitters with a strong transition dipole and to confine the electromagnetic field in a small volume.

The coherence of the electromagnetic field in the dressed states makes it possible to couple coherently a collection of $N$ oscillators in a cavity. This improves the effective coupling strength: $g_N\sim\sqrt{N}g$. In the end, the coupling strength  $g_N$ is then proportional to the oscillator density in the cavity (because $g$ is inversely proportional to $\sqrt{V_c}$). Thanks to collective effects, the strong coupling regime becomes accessible at room temperature. In the case of $N$ oscillators coupled to one cavity mode,  $N+1$ polariton states emerge. However, two of them only are bright states, while the other $N-1$ states are unobserved dark states; thus the situation remains similar to the case of a single two-level system (see for example the review \cite{AccountsChem49-2016-2403-Ebbesen}).

\subsubsection{Quantum models}

Notwithstanding the rapid progress in modeling polaritons, more research is still crucially needed to fully understand their properties. In a very original direction, the concept of polariton was integrated into an approach of density functional theory (DFT) that takes into account the existence of a cavity when evaluating the structure and the physico-chemical properties of molecules \cite{PNAS114-2017-3026-Flick}. Both matter and the quantum field are modeled in detail, combining two research fields, within material and electromagnetic sciences, in order to extract precise information on the nature of the states and their photonic, vibrational, and electronic components.

Other research directions consist in the study of the influence of the ultra-strong coupling regime (as, for example \cite{JPCC119-2015-29132-GarciaVidal, PRA93-2016-033840-Cwik}  aimed at elucidating the enhancement mechanisms in Raman scattering). The ultra-strong regime is defined by a Rabi frequency that becomes similar to the cavity mode frequency ($\Omega_0\approx\omega_c$). In the ultra-strong coupling regime, the rotating wave approximation (RWA) does not hold. As a result, even the ground state of the quantum system is modified by the coupling. (For example, the ultra-strong coupling regime was reached experimentally for vibrational polaritons in molecular liquids \cite{PRL117-2016-153601-Ebbesen}). In parallel, groups pursue the investigation of the influence of the coupling between the different degrees of freedom in polaritonic systems. It is important to consider the vibronic structure to model fluorescence \cite{CPL683-2017-653-Mukamel}, while the effects of anharmonicity of the vibrational modes are also investigated \cite{JCP144-2016-124115-Mukamel} under the strong coupling regime. The study of the interaction between photons, excitons, and vibrations is useful for understanding luminescence and light-harvesting complexes \cite{PRB94-2016-195409-GarciaVidal}. Advanced models predict more subtle effects such as differences between the emission and aborption rates from polaritons or contributions from polaritonic dark states \cite{PRA95-2017-053867-Herrera}. Progress can also come from the optical modeling side, such as this approach inspired from transformational optics \cite{PRL117-2016-107401-GarciaVidal}.  The problematics of quantum decoherence is also very important. The influence of the dephasing processes of individual molecules on the collective strong coupling regime was studied for molecules coupled to a common bath (long-range phonons) and to indivual baths (short-range phonons) \cite{NJP17-2015-053040-delPino}. In the case of the common bath, it was found that polaritonic dark and bright states remain uncoupled. In the case of individuals baths, incoherent energy transfers may occur between bright and dark states. However, these processes can be neglected in practice if the coupling $\Omega$ is larger than the frequencies of the phonon bath. This study confirms the usefulness of these systems for studying opto-mechanical strong couplings at room temperature.

In the following two sections, we review briefly the basics of photonics and plasmonics, accompanied with a description of the structures used to achieve strong coupling.

\section{Photonic-dielectric structures for SC}

Photonic crystals (PCs) represent a broad range of dielectric periodic(see the figure 5) and quasi periodic structures, ideally suited to create hybrid light-matter states \cite{Kolaric},\cite{Lando}. PCs are designed to control and manipulate not only photon propagation, but more importantly also, photon generation.

\begin{figure}\label{fig:Photonic crystals}
\includegraphics[width=7cm]{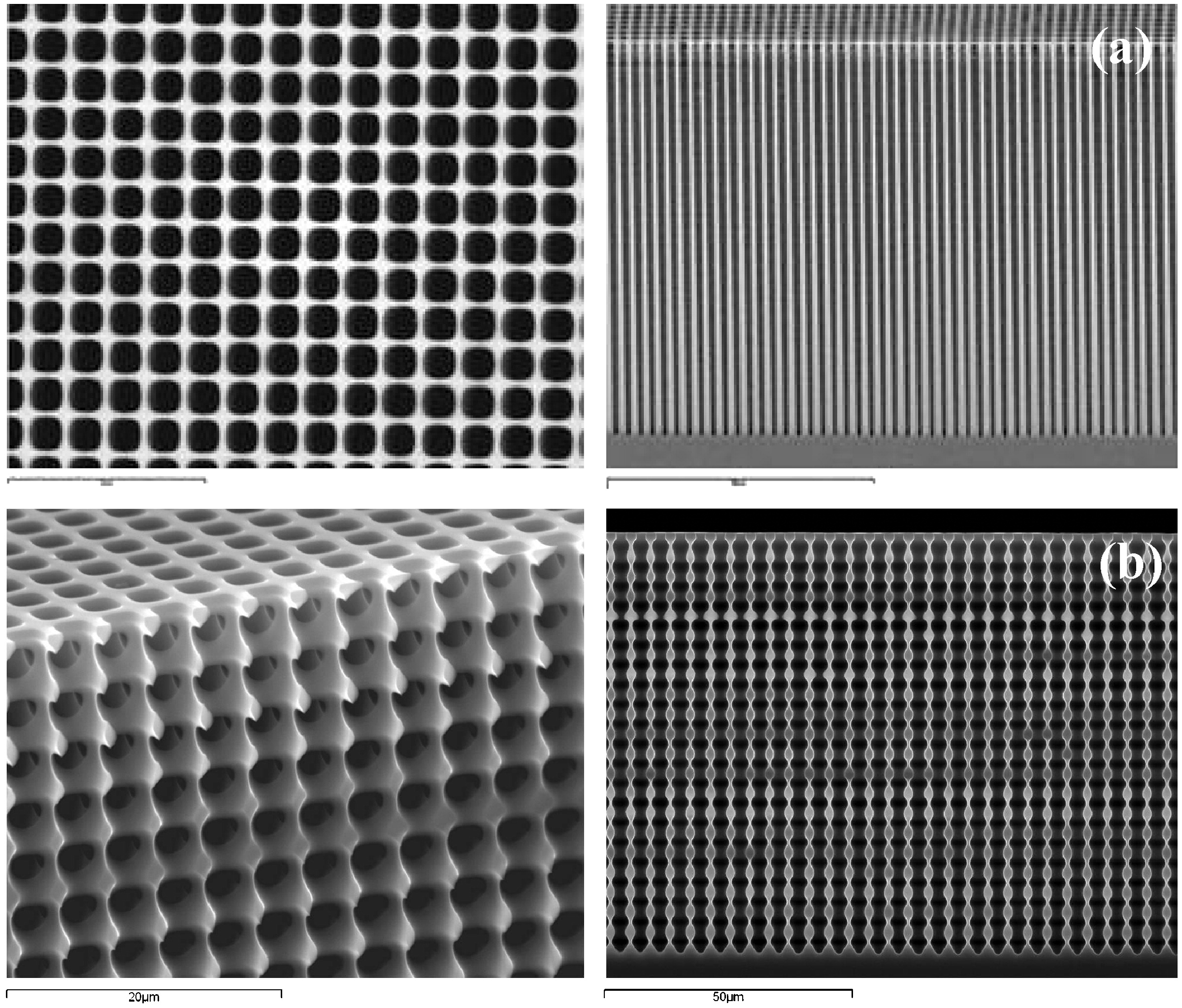}
\caption{\footnotesize \itshape SEM pictures of various  (a) 2D and(b) 3D macro-porous silicon  structures for various photonic applications, taken from ref.[7]}
\label{one}
\end{figure}

\pagebreak

 In the first approximation PCs can be imagined as the photonic equivalent of electronic semiconductors, which are also widely used to manipulate the flow of electrons \cite{Prasad,Joannopoulos}. The photonic bandgap (PBG) is the photonic equivalent of the electronic bandgap in semiconductors that defines a range of forbidden energies for the electrons between the valence and the conduction band, while PBG defines the range of forbidden wavelengths (or, equivalently also energies) for photons.  This spectral range of forbidden wavelengths for the dielectric periodic structure is intuitively clear from the notion of destructive and constructive interference between light waves diffracted from a periodic structure with a periodicity in the optical wavelength range. 
The Bragg relation  $2 d_{hkl} \sin\Theta = n_e \lambda$, usually used for Bragg diffraction of X-rays for incidence angle($\Theta$) from atomic, ionic or molecular crystals, relates the periodicity $d_{hkl}$ of the structure represented by the distance between consecutive lattice planes with Miller indices $(h,k,l)$ to the wavelength $\lambda$ of the light diffracted by the structure with effective refractive index $n_e$, hence not propagating in the structure. The width of the PBG is inversely proportional to the dielectric contrast. Introducing defects in the periodicity also introduces defects in the forbidden PBG, creating additionally allowed states, which have been termed p- or n-doped according to their spectral position relative to the lower or higher energy wavelengths. This concept of forbidden stopband with allowed defect mode pass bands allows the description of fluorescence suppression when the photon is emitted in the forbidden stopband, and spectral narrowing, when the emission is confined in the allowed pass band. The forbidden PBG can also be described in terms of its Density of Optical States (DOS), the number of electric field modes per unit volume at a given frequency (or wavelength) to which the transition dipole moment for emission can couple. For the DOS in a PC, the local DOS (LDOS) is calculated from the eigenfrequencies of the eigenmodes of the electric field in the PC and an integration is performed over all wave vectors within the first Brillouin zone and a summation over all band indices. Following Fermi's Golden Rule, the rate of spontaneous emission is proportional to this LDOS. Time-resolved experiments then allow the comparison between the calculated (lowering of the) LDOS and the (decreased) photon emission rate. 

Different approaches towards engineering the dielectric periodicity are being used. There is a trade-off between convenience of fabrication and extent of the control over the emission. Colloidal photonic crystals are fabricated by inexpensive, chemical bottom-up self-assembly techniques. Drop-casting, convective self-assembly, spin-coating of colloidal particles, mainly highly monodisperse and spherical polystyrene or silica particles (photonic atoms) provide crystals with a photonic bandgap depending on the size of the particles used.  Doping can be realized by integrating a single layer of a particle with different size or different refractive index (or both, only an optical phase shift is needed) by depositing a monolayer by the Langmuir-Blodgett (LB) technique.  Such direct opals have the disadvantage of a low refractive index contrast ($n_{PS} \approx 1.6$, $n_{Silica} \approx 1.4$, $n_{Air} = 1$), and a large fill factor for hexagonal closed packing of 0.74. Therefore, the bandgap is only a pseudo-bandgap, in contrast to the full bandgap, which is independent of incidence angle of the light. The influence on the emission intensity and the emission rate of an inserted molecular fluorophore is clearly there, as has been witnessed by enhanced energy transfer from a donor molecule with its emission suppressed in the bandgap \cite{Kolaric1,Vallee}; and by spectral narrowing of the emission suppressed by the broad forbidden stopband, yet enhanced in the allowed passband. But as a token of the relatively small effects, the spectral narrowing, as a necessary, yet not sufficient, condition for lasing, did not result in lasing in the photonic heterostructure\cite{KoenACS}. Also quantitatively, the decrease in emission rate was small (3 to 4 \%), yet in good agreement with the calculated LDOS.

One strategy to enhance the refractive index contrast and lower the fill factor, is inverting the structure by using the direct opal as a template for inversion by filling the air voids with a higher refractive index material, e.g.\ titania, followed by removing the templating material (etching silica or calcining polystyrene). Stronger influences on the emission rate of quantum dots (to 40\%) could be observed \cite{Lodahl}.
Another strategy is based on hollow silica spheres, combining the convenience of direct self-assembly of colloidal particles with the large refractive index contrast with the inserted interstitial medium. Together with the improved quality of the transfer of a monolayer by lifting from a free floating layer, compressed by surfactant molecules acting as soft barriers (as opposed to the hard barriers in LB deposition) this has allowed defect-mode passband lasing in a self-assembled photonic crystal laser.  This has been attributed to the relative narrowness of the allowed pass band, resulting in a quality factor Q of about 50. 
Although a clear splitting by the photonic bandgap of the emission band in a lower and a higher energy band can be observed, the effect is still in the weak coupling regime, described as a Purcell effect. 

Examples of strong coupling have been achieved starting from solid absorbing material in a bulk or thin film format in a microcavity composed of thin reflecting metal  layers.  These are the equivalent of discrete mirrors forming a cavity comprising the active absorbing and/or emitting photonic material.  Photonic crystals provide distributed feedback and are mainly comprised of passive dielectric material.  The active materials added so far have been limited  to emitters (molecular fluorophores or quantum dots). As the emitted photons are detected with very high efficiency, very few emitters can be detected, but this does not lead to the large oscillator strength needed for the strong coupling regime \cite{Cohen}.
An avenue that could be pursued is the combination of the dielectric periodicity providing the distributed feedback and the strongly absorbing material in an inverted opal made from that material itself, or, based on the hollow spheres, to fill the interstitial voids completely by that material.

\section{Plasmonic-metallic structures for SC}

Plasmonic micro- and nanostructures have become an important paradigm in photonics for fundamental experiments and applications. Currently, metallic and hybrid metallo-dielectric devices are examined for sensors, solar cells, light emitting diodes (LEDs), imaging, nonlinear effects, metasurfaces and metamaterials, medical applications, nanocavities, nanocircuits and nanolasers, higher-order effects and so on \cite{Atwater2010,Cai2010,Oulton2009,Rivera2016}.
One of the outstanding features of plasmonics versus dielectrics is the access to very high values of the wavevector, leading to very small effective wavelengths, much smaller than the typical diffraction limit \cite{Maier2007}. This makes it possible to confine the light in extremely small volumes, in turn enhancing a range of phenomena that are inaccessible with purely dielectric structures. 

The confinement leads to more intense electromagnetic fields in a certain volume of matter, which can be used in myriad ways: enhanced absorption (for solar cells, local heating), emission (LEDs), gain (lasers) and so forth. The physical origin of the plasmonic confinement stems from the coupling between light and electronic charge oscillations on a metallic interface. Fundamentally, these excitations are thus a kind of quasi-particle, which are called surface plasmon polaritons (SPPs), which is typically shortened to plasmons. The standard materials for plasmons are noble metals such as gold and silver for visible and near-infrared modes. However, the opportunities are much wider, with now an extensive range of materials yielding plasmonic modes: doped semiconductors, oxides, two-dimensional materials such as graphene (for the mid-infrared) etc. \cite{Tassin2012}

In principle, for strong coupling any kind of optical resonance can be used, but the plasmonic modes offer a very interesting family of resonances, with strong localization along one, two or three dimensions. The most basic geometries are the planar metal-dielectric surface, and the metallic particle, respectively. However, many other geometries, structures and combinations are investigated, with often unique and tailorable features. For example, the effect of SC on quantum emitters embedded within a plasmonic bowtie structure is presented in Figure 6.

\begin{figure}\label{fig:bowtie}
\includegraphics[width=7cm]{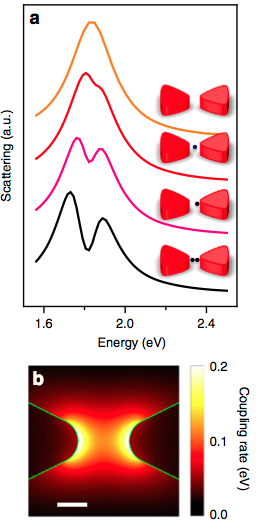}
\caption{\footnotesize \itshape Electromagnetic response of strongly coupled plasmonic (bowtie) structures. Simulated scattering spectra of bowties with one or two quantum dots as indicated in the inset structures. (b) Distribution of the coupling rate (in eV) of a quantum emitter with an oscillator strength of 0.6 in a bowtie structure. The white bar represents 10 nm, taken from ref. [79].}
\label{one}
\end{figure}

\pagebreak

As explained in Section \ref{sectionApplications}, strong coupling has a renewed interest because it can influence a large variety of phenomena, which at first sight have no connection with optics. For some effects, such as these involving chemical reactions, it is interesting to have direct access to the optical mode, which leads to the concept of an open cavity. Indeed, the traditional cavity for strong coupling is a straightforward Fabry-Perot type resonance with two mirrors, so a closed cavity, even though there are variations possible \cite{Flatten2016}. The material component must therefore be sandwiched in between the metallic or dielectric interfaces, which is not always easy to accomplish in practice. Alternatively, there are plasmonic architectures that allow for a directly available surface, so that many different materials (molecules, two-dimensional materials, nanoparticles etc.) can be integrated in a flexible way. For strongly coupled chemical reactions for example, a microfluidic structure on top would prove a useful testing ground, so that the optical and material resonances are more or less decoupled.

The simplest plasmonic open resonance consists of coupling with a surface plasmon mode, via a Kretschmann type configuration. In this way a surface mode is excited via a higher-index prism on the bottom. As the evanescent mode is excited on the other surface of the metal layer, any material deposited on the metal will be in close contact with the mode. This geometry was employed for electronic \cite{Pockrand1982,Bellessa2004} and recently also for vibrational strong coupling \cite{PRL117-2017-126802-Memmi}.

Another, much more tailorable architecture is offered by a regular array of metallic nanoparticles. With the correct dimensions these structures can exhibit a kind of hybrid dielectric-plasmonic mode, the surface lattice resonance (SLR) \cite{Kravets2008}. These SLRs couple a localized plasmon with a diffractive mode in the plane (connected to so-called Rayleigh anomalies). This mixing can lead to resonances with relatively high quality factors, and tailorable confinement along the plane of the particles \cite{Abass2014}. Therefore when these particles or nanorods are fabricated on a surface, any material deposited on top can be in close contact with the optical mode, thus one can obtain and optimize an open cavity. 

Strong coupling has been demonstrated in a wide range of plasmonic cavities and materials, we refer to \cite{RPP78-2015-013901-Barnes,Marquier2017,Baranov2017}for recent excellent reviews. Various phenomena with strongly coupled plasmonic lattices were recently reported \cite{Vakevainen2013,Rodriguez2013,Ramezani2017}. These optical modes are stretched over a sizeable distance in the plane, hence coupling with a significant amount of absorbing molecules. Therefore, there is another line of research towards full three-dimensional subwavelength confinement of the optical mode, with for example coupled particles, or the highly effective particle on a surface setup. The latter design has recently been used for single molecule strong coupling, so with a single molecule in the resonant field of a single nanoparticle mode \cite{Chikkaraddy2016}.

An exciting development in this context is the realization that quenching can be controlled \cite{Faggiani2015,Li2016}. When an emitter is put close to a metallic element, typically within a few nm or below, a rapidly increasing loss is observed. This non-radiative phenomenon is called quenching, and limits the radiative efficiency of many processes in plasmonics. However, it was established that quenching stems from coupling to higher-order (less or non-radiative) modes in the devices. Moreover, within appropriate geometries, this coupling can be reduced drastically, or the higher-order modes can become more radiative. For various plasmonic structures, this explains the large radiative efficiencies that are being achieved, beyond the capabilities of dielectric structures. Controlled quenching is also the origin for single molecule strong coupling reported with the nanoparticle on a surface geometry \cite{Kongsuwan2017}.

\section{Strong coupling applications}\label{sectionApplications}
 
\subsection{Entanglement and information}

The strong coupling regime generally requires cavities with high-quality factors $Q_c=\omega_c/\gamma_c$ but it is also possible to exploit surface plasmon polaritons (SPP) \cite{RPP78-2015-013901-Barnes} thanks to their strongly localized evanescent electric field, which provides a high density of states in a small volume \cite{NanoLett17-2017-3246-Qiu}. This approach has made possible the coupling of excitons with nanoparticle SPP \cite{NanoLett17-2017-3246-Qiu} see also the previous section , as well as of molecular vibrations with a metallic film SPP \cite{PRL117-2017-126802-Memmi}. Although quantum strong coupling has a classical analogue \cite{ACSphot4-2017-1669-Stete}, the purely quantum nature of the polariton states makes them useful for quantum information processing. Several experiments demonstrated that SPP preserve quantum information and entanglement. For example, they transmit the spin-orbit coupling information of incident light and allow the post-selection of the final state in a quantum weak measurement of the light chirality \cite{PRL109-2012-013901-Ebbesen}. They preserve polarisation entanglement of photon pairs \cite{Nat448-2002-304-Altewischer} as well as energy-time entanglement \cite{PRL94-2005-110501-Gisin} in cascading photon to SPP to photon conversions, with a preservation of temporal coherence that is larger than the SPP lifetime  \cite{PRL94-2005-110501-Gisin}. In addition, both the quasi-particle and wave nature of SPP were highlighted in experiments similar to those attesting the wave-particle duality  of photons when they interact with beamsplitters \cite{NatPhys5-2009-470-Kolesov}. 

Providing evidence of the corresponding properties for cavity polaritons is an inviting research prospect given that the strong coupling of quantum oscillators is a truly needed tool for quantum information processing.
 
\subsection{Vibrational polaritons}
Polaritons originating from the coupling of a vibration to a cavity mode redefine the very structure of matter. Changing the oscillator strengths of molecular vibrations modifies bond lengths and the energy surfaces associated to bond formation and dissociation. By reaching the strong coupling regime in a microfluidic cell \cite{PRL117-2016-153601-Ebbesen}, it is even possible to alter the rate of chemical reaction \cite{Angew55-2016-11462-Ebbesen}. Here, chemical reactivity is controlled differently than in catalysis or photocatalysis, since the rates are modified even in the absence of light! This is catalysis \textit{ab vacuo}. 
The strong coupling regime is also at the origin of phase transitions in optomechanical cavities \cite{PRL119-2017-043604-Cortese}, comparable to the alignment of dipoles in a polymer. Strong coupling has thus fundamental macroscopic consequences.

\subsection{Spectroscopic properties}

Strong coupling modifies substantially the spectroscopic properties of molecules and their optical response (the intensities and frequencies of infrared absorption, Raman scattering, nonlinear emission such as second-harmonic generation SHG, and fluorescence). Vibrational polaritons were evidenced by Fourier transform infrared spectroscopy (FTIR) and were probed as a function of the IR incidence angle. Several groups determined the dispersion law of stongly coupled stretching vibrations of CO groups in polymers (polyvinyl acetate PVAc or poly methyl methacrylate PMMA) placed inside metallic \cite{NatCom6-2015-5981-Ebbesen, ACSPhot2-2015-130-Simpkins} or dielectric (Dielectric Bragg Reflector - DBR)  \cite{AnnPhys528-2016-313-Tischler, JPPCL7-2016-2002-Tischler} cavities.

In the figure 7 three dispersion branches are presented  as a consequence of the formation of  polaritons  of distinct CO groups from a PMMA polymer matrix and dimethylformamide molecules in SC regime \cite{JPPCL7-2016-2002-Tischler}. Presented dispersion are extracted from angular resolved transmission spectra.

\begin{figure}\label{figure1}
\includegraphics[width=\columnwidth]{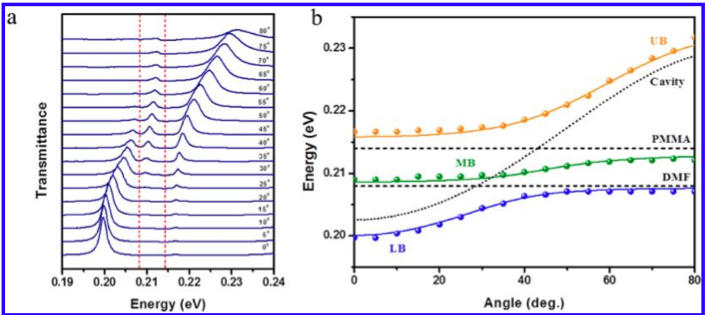}
\caption{\footnotesize \itshape (a) Room temperature angle resolved transmission spectra . b) The vertical red dashed lines mark the vibrational energies of the carbonyl stretches in the DMF (EDMF = 0.208 eV) and PMMA (EPMMA = 0.214 eV). Dispersion relations of the microcavity vibrational polartion states (b) extracted from angle-resolved transmission spectra. The circles are the polariton energies obtained from the transmission peaks; orange, green, and blue represent upper, middle, and lower branches of the dispersion relation, respectively. The solid curves are the theoretical fits using a three-level coupled oscillator model. Broken curves correspond to the dispersion relations of the uncoupled photon (dotted curve) and vibrational (horizontal dashed lines) resonances, taken from ref.[67]}
\label{one}
\end{figure}

The transition from weak to strong coupling was investigated for the polaritons of the stretching vibration of the isocyanate NCO group \cite{ACSphot2-2015-1460-Simpkins}. Strong coupling of  $\textrm{COO}^-$ bonds in biological molecules has been achieved, opening the way towards the control of enzymatic reactions \cite{JPCL7-2016-4159-Ebbesen}. 

The Raman scattering cross-section of the stretching vibration of CO groups in PVAc polymer exhibits a two-to-three orders of magnitude enhancement under strong coupling conditions  \cite{Angew54-2015-7971-Ebbesen} (which is not due to a simple local field enhancement in the cavity). A model of Raman scattering under strong coupling conditions points out that the total Raman cross-section is expected to remain constant while it is divided amongst the two polaritons \cite{JPCC119-2015-29132-GarciaVidal}. The enhancement could thus originate from a modification of the ground state itself, an effect normally associated to the ultra-strong coupling regime (where the rotating wave approximation RWA fails) \cite{JPCC119-2015-29132-GarciaVidal}. A similar enhancement occurs for the nonlinear SHG signal from c-porphyryn molecules placed inside an Ag cavity \cite{NanoLett16-2016-7532-Ebbesen}. In another nonlinear experiment, the interference of the non-resonant SHG signal with the resonant SHG signal coming from the exciton-polariton of ZnO micro-wires gives rise to a Fano resonance \cite{PRL118-2017-063602-Wang}. The latter is a direct indication of the coherence of the polaritons.

\subsection{Fundamental aspects and nonlinear effects}
Furthermore, polaritons exhibit fundamental quantum properties, such as Bose-Einstein condensation. The condensate properties (coherence, polarisation, population) inside the cavity can be probed from outside the cavity thanks to the far-field component of the polaritonic states \cite{Nat443-2006-409-Kasprzak}. A proposal has been made to use polaritons as the fluid in a quantum thermal machine based on the Casimir effect between two nano-resonators  \cite{PRE95-2017-022135-Tercas}. Polaritons are also considered for the development of treshold-less lasers (the one-atom laser is made possible because the atom can only emit in the cavity single mode) \cite{MicroCav-book-Kavokin}. Reaching the strong coupling regime for single oscillators, such as in this case with an an electronic transition at low temperature \cite{PRX7-2017-021014-Wang}, lets us glimpse their future use for nonlinear effects; indeed, the measurement in these systems of a second-order correlation $g^{(2)}$ near $0$ indicates the possibility of making a single photon source from their emission. Bow-tie plasmonic cavities also reached the strong coupling regime for a single quantum dot \cite{NatComm-2016-11823-Santhosh}. Similarly, a nano-plasmonic cavity was strongly coupled to a single molecule at room temperature \cite{Nat535-2016-127-Baumberg}. Research on the strong coupling regime for single emitters are useful to develop quantum nonlinear optics in the photon by photon regime (which could be used for example to develop non-destructive photon counting detectors \cite{NatPhot8-2014-685-Lukin}). Indeed, the optical reponse of a two-level system is fundamentally nonlinear since a second photon does interact with a system state that was modified by its interaction with the first photon (for a review, see  \cite{NatPhot8-2014-685-Lukin}). A single photon interacting with a single atom strongly coupled to a cavity can block the cavity transmission, while two photons can make the system climb the Jaynes-Cummings ladder.

\subsection{Coherence, delocalisation and transport properties}

The coherence of energy transfer in materials plays an important role in defining their functionality, as for example in the biological processes of light-harvesting \cite{Nat543-2017-647-Scholes}. The quantum coherence of polaritons gives opportunities to control transport properties of materials. For instance, the study of the energy transfers inside matter under the strong coupling regime by two-dimensional infrared spectroscopy \cite{JCP144-2016-124115-Mukamel} for the amide-I (CO) and amide-II (CN) vibrations demonstrated that the coupling influences directly the vibrational excitation lifetimes. The vibration lifetimes are also strongly modified in this pump-probe infrared absorption study of the CO group stretching vibration \cite{NatCom-2016-13504-Dunkelberger}, with additional evidence of coherent energy transfers between the two polaritons provided by the detection of quantum beats.
The observation of energy transfers between donor-acceptor molecules separated by more than 100 nm, much further apart than normal transfer distances, evidences the delocalized and coherent nature of polaritons \cite{Angew56-2017-9034-Ebbesen}. Polaritons can thus entangle cyanine dyes at a distance. Studies of the transient absorption dynamics of donor-acceptor systems also showed the modification of the non-radiative transfers in the cavity  \cite{Angew55-2016-6202-Ebbesen}.
Morever, strong coupling can transform a non-radiative deexcitation pathway into a new emission line in the cavity, thereby transforming a Raman laser into an optical parametric oscillator (OPO) \cite{PRL117-2016-277401-GarciaVidal}.
Strong coupling modifies also the energy exchanges associated with vibronic states in the fluorescence process, giving rise to superradiance or subradiance for collective emissions, with polariton-dependent coupling strengths \cite{CPL683-2017-653-Mukamel}.
Strong coupling can thus activate or deactivate energy levels in the energy transfer pathways and it affects substantially the transport properties as a result of its coherence.

Remarkably strong coupling was used to mold electrical conductivity of semiconductor films within metallic cavities. Recently the Ebbesen group \cite{Orgiu} showed that strong electronic coupling between organic thin film transistors and plasmonic modes of the open metallic cavity creates coherent dressed states that extend over $10^5$ molecules with a significantly different conductivity. The presented research thus offers a unique way to mold material transport properties, such as conductivity, without changing the chemistry of the materials in a classical way.

\subsection{Excitation by quantum sources}
The quantum properties of polaritons also play also a significant role when polaritons are excited by quantum light sources, to create then purely non-classical states \cite{PRL115-2015-196402-Lopez}. The schematic view of quantum optical experiment with polariton using quantum light is presented in Figure 8 \cite{PRL115-2015-196402-Lopez}.

\begin{figure}\label{figure1}
\includegraphics[width=\columnwidth]{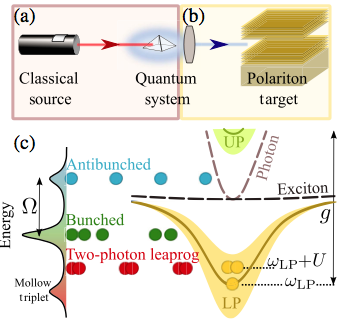}
\caption{\footnotesize \itshape Scheme of optical experiment with cavity polaritons using quantum light, (a) A typical optical excitation scenario of quantum optics, a laser excites a quantum system, e.g., a quantum dot. (b) Instead of the conventional scenario of also exciting polaritons with a laser, we excite polaritons with quantum light: specifically, from the output of the quantum system excited by the laser. (c) Mollow spectroscopy: the photoluminescence of a strongly driven two-level system provides the Mollow triplet, shown on the left with energy on the vertical axis. Various spectral windows provide different types of photon correlations, sketched here as photon balls with different temporal spacing. Exciting the lower polariton (LP) with leapfrog photon pairs allows us to measure accurately very small values of the interaction, taken from ref. [86]}
\label{one}
\end{figure}

\pagebreak

Indeed, the Jaynes-Cummings model shows that these(non-classical) states can be generated using non classical sources \cite{NatPhot8-2014-685-Lukin} (thermal or laser light are considered classical while one or several photon sources are considered quantum). The study of the excitation polaritons by these sources are under progress (see for example this theoretical study of the coupled harmonic oscillator as well as of the coupled two-level system \cite{PRA94-2016-063825-Carreno,PRA94-2016-063826-Carreno}). Polaritons can also encode the orbital angular momentum of light and can control of the spin-orbit states for quantum information applications  \cite{PRB94-2016-201301-Li}.

\subsection{Phase transitions}
Recently it was shown that the concept of SC could be extended in the classical realm of the physics of phase transitions\cite{Nanoscale}. Introducing a perovskite inside a metallic cavity allows the formation of hybrid states with a significantly different local structure, i.e., ground states, from the noncoupled counterpart. The latter strongly affects the material phase transitions, so the collective behavior of atomic entities. Thermodynamic studies show that the ground state of strongly coupled molecules is lowered as the Rabi splitting increases. However, SC is also associated with splitting excited states, which also holds for the ground state energy landscape such as demonstrated in the case of a phase transition. Increasing the Rabi splitting enhances the phase transition dynamics, first by the direct perturbation of the ground state and secondly, by increasing the fraction of coupled oscillators in the cavity. These findings are extremely significant since they offer possibilities to control collective (thermodynamic) properties of matter and self-assembly processes, by controlling the interaction of the vacuum field inside the cavity with a molecular entity (molecular or atomic oscillators). 

\subsection{SC in biology}

In recent work, Vedral extends the concept of SC to biology and shows that the concept \cite{Vedral1,Vedral2} can be used to entangle macroscopic bacteria with the field. He presented experimental proof where living sulfur bacteria are entangled with a quantized light field. Presence of Rabi splitting of two coupled harmonic oscillators (light and bacteria, respectively) confirm that they are entangled. Specifically, the entangled subsystems are the excitons in the bacteria and the photons in the cavity\cite{Vedral1}. The remarkable fact is that the bacteria stay alive during the experiments, despite the fact that they are strongly coupled to the cavity light. Additionally, a critical element relating to the nature of entanglement between matter and field was pointed out: Since light is known to be quantum mechanical, and the Rabi splitting is observed, the bacteria (more precisely, whatever degree of freedom within the bacteria that couples to light) also has to be a quantum system, regardless its macroscopic size \cite{Vedral1},\cite{Vedral2}. The  SC concept can thus be used to achieve macroscopic quantum coherence at ambient conditions within wet and warm biological matter. These new studies will have a significant impact in the fields of quantum biology and the physics of open quantum systems.

\section {Conclusion and highlights}

In this review, we presented the theoretical background of the SC concept and its conceivably unlimited applications in the field of material science. The essential element of SC is entanglement between matter and the light field, which has no classical analog and significantly affects the properties of matter. We discussed in detail the importance of the concept of entanglement, especially between matter and the electromagnetic field, for chemical and material applications.

SC profoundly connects material science with fundamental physics and quantum information science, and offers the possibility to control and mold material and molecular properties (phase transitions, reactivity, spectroscopy, etc.) through the hybridization with the quantum field (quasi-particle design). The general application of this concept could lead to a new scientific revolution, and could be applied to engineer novel complex materials, such as high-temperature superconductors, topological materials, semiconductors, etc. It can be employed to control biological dynamics and to affect chemical reactivity in an unprecedented way, just by using and harvesting interactions between the vacuum field and material oscillators. Additionally, SC offers a quantum mechanical control on classical processes, such as phase transitions and self-assembly, creating a remarkable link between the quantum and classical realm. Furthermore, engineering structures for SC could rapidly lead to new breakthroughs in the field of nanoscale optics and photonics. Regardless of the fact that, at the moment, SC is still at its infant stage of development, the primary aim of this article is to attract the interest of the broad chemical community for the SC concept, which could be used to open new horizons in chemical and material research.
In the end, it should be stressed that, in life but especially in science, we often have a trade-off between beauty and practicality, between dreams and reality. The SC concept unifies these things, being a magnificent dream inside the beautiful dream which we call physics.

\section{Conflicts of interest}
There are no conflicts of interest to declare.

\section{Acknowledgments}
Y.C.\ is a research associate of the Belgian Fund for Scientific Research F.R.S.-FNRS. B.K.\ and B.M.\ acknowledge support from the Interuniversity Attraction Pole: Photonics@be (P7-35, Belgian Science Policy Office), and from the F.R.S.-FNRS. T.D.\ warmly acknowledges the contribution of V.\ Debierre and I.\ Goessens to the numerical simulations \cite{goessens} in relation with the discrete Fermi Golden Rule. 
 B.K. and T.D. acknowledge support of the COST Action 1403 Quantum Nano Optics $http://www.cost.eu/COST_Actions/mpns/MP1403$.
 Additionally B.K.\ warmly acknowledge the fruitful and inspiring discussion with Dr.\ Ana Cvetkovic (FAR Polymers s.r.l., Milano, Italy) considering applications of SC concept in chemical science.

\end{document}